\documentclass[fleqn,twoside]{article}%
\usepackage{amsmath}
\usepackage{amsfonts}
\usepackage{amssymb}
\usepackage{graphicx}
\usepackage{cite}
\def\be{\begin{equation}}

\def\ee{\end{equation}}

\def\bi{\bibitem}

\begin{document}

\title{Irrotational Bianchi V Viscous Fluid Cosmology with Heat Flux.}

\author{A. Banerjee and Abhik Kumar Sanyal}
\maketitle

\noindent

\begin{center}
Department of Physics, Jadavpur University, Calcutta-700032, India  \\
\end{center}

\begin{abstract}
\noindent The irrotational Bianchi V cosmological model under the influence of both shear and bulk viscosity, together with heat flux, has been studied. Exact solutions for the model are obtained with three physically viable assumptions. The first two relate the matter density, shear scalar, and expansion scalar and the third is a barotropic equation of state, connecting the matter density and thermodynamic pressure. The properties of the solutions are studied and the temperature distribution is also presented explicitly. It has been observed that along with the viscosity, heat flux further adds to the rate of entropy increase.
\end{abstract}

\section{Introduction:}

When the constituent fluid in a cosmological model exhibits dissipative phenomena such as viscosity, heat flow, etc., the energy momentum tensor for a perfect fluid is no longer valid. Additional terms due to these effects are to be included and the solutions obtained are different from those known for perfect fluid cosmologies. The consideration of viscosity and heat flow may be important for two reasons. Both the currently observed large entropy per baryon and the remarkable isotropy of the background radiation in the present epoch can successfully be explained \cite{1,2}. The initial singularity is also of a different kind when the irreversible processes are considered. \\

Exact solutions corresponding to such dissipative fluids in cosmological models are worth investigating from the above considerations.
Previously many workers \cite{3, 4, 5, 6, 7, 8} obtained exact solutions corresponding to Bianchi $\mathrm{I, II, III, VI_0}$ and Kantowski-Sachs spacetimes with viscous fluid, and in some cases with the addition of an axial magnetic field. However, within best of our knowledge, heat flux has not been handled in the background of anisotropic cosmological models, so far. Nevertheless, dissipative phenomena due to the inclusion of heat flux has been discussed previously Bergmann \cite{9}, and also by Glass \cite{10}. Bergmann \cite{9} obtained a generalized Robertson-Walker metric for a shear-free perfect fluid with a radial heat flow vector and Glass \cite{10}, using almost the same technique, obtained exact shear-free collapse with heat flow in radial direction for a bounded spherical fluid. Later, Sanyal and Ray \cite{11} presented cosmological solutions of
Einstein's equation with heat flow, generalizing the results of Bergmann. Recently Modak \cite{12} considered some spherically symmetric cosmological solutions with heat flux in the radial direction. The velocity vector of the fluid is shear-free, and thus the new set of solutions obtained in this paper includes those of Bergmann \cite{9} and Maiti \cite{13} as special cases. As mentioned, the anisotropic cosmological models of a Bianchi group admitting heat flux are, however, not quite available in the literature.\\

In the present paper a Bianchi V model corresponding to a viscous fluid together with heat flow is considered. Since the number of unknown functions exceeds the number of field equations by three, to obtain solutions we have assumed three additional restrictions. Two of these are a pair of linear relations among the matter density $(p)$, shear scalar $(\sigma^2)$, and expansion scalar $(\theta)$, and the third one is a barotropic equation of state. The singularity is found to exist in finite past in such models. The temperature distribution is explicitly obtained when the thermal conductivity is a function of time alone because of the homogeneity consideration. It is found that the temperature gradient, along with the other scalars such as $\rho,~ \sigma,~ \theta$, etc., attain infinite magnitudes at the initial singularity of an expanding model.

\section{Einstein's Field Equations and General Expressions:}

The metric for irrotational Bianchi V cosmological model is,

\be ds^2 = - dt^2 + e^{2\alpha}dx^2 + e^{2x}(e^{2\beta} dy^2 + e^{2\gamma}dz^2),\ee
where, $\alpha$, $\beta$ and $\gamma$ are functions of time alone. The proper volume is given by

\be R^3 = \sqrt{-g} = e^{\alpha + \beta + \gamma + 2x},\ee
where $g$ is the determinant of the metric (1). The energy-momentum tensor under the influence of viscosity and heat flux is

\be \begin{split} &T_{\mu\nu} = (\rho + \bar{p}) v_\mu v_\nu - \bar{p} g_{\mu\nu}- \eta U_{\mu\nu} + q_\mu v_\nu + q_\nu v_\mu,\\&
\mathrm{where,}~~\bar p = p - \left(\zeta - {2\over 3}\eta\right)v^\alpha_{~;\alpha},~~\mathrm{and}~~U_{\mu\nu} = v_{\mu;\nu} + v_{\nu;\mu} + v_\mu v^\alpha v_{\nu;\alpha} + v_\nu v^\alpha v_{\mu;\alpha}.\end{split}\ee
In the above, $p$ and $\rho$ are the isotropic pressure and matter density of the fluid respectively, $\bar p$ is the effective pressure, $\eta$ and $\zeta$ are the shear and bulk viscosity coefficients respectively, $v_\mu$ and $q_\mu$ are the four-velocity and heat flow vectors, respectively, so that

\be v_\mu v^\mu = -1, ~~~\mathrm{and}~~~q_\mu v^\mu = 0.\ee

In the system of units $8\pi G = c = 1$, the existing components of Einstein's field equations

\be R_\mu^\nu - {1\over 2} \delta_\mu^\nu R = - T_\mu^\nu,\ee
in the co-moving coordinate system, $v^\mu = \delta_0^\mu$, and $q^\mu = \delta_1 ^\mu q^1= 0$, are the following,

\be \dot \alpha \dot \beta + \dot \beta\dot \gamma + \dot \gamma\dot \alpha - 3e^{-2\alpha} = \rho,\ee
\be \ddot\gamma + \dot \gamma^2 + \ddot\beta + \dot\beta^2 + \dot\gamma \dot\beta - e^{-2\alpha} = -\bar p + 2\eta \dot \alpha,\ee
\be \ddot\beta + \dot \beta^2 + \ddot\alpha + \dot\alpha^2 + \dot\beta\dot\alpha - e^{-2\alpha} = -\bar p + 2\eta \dot\gamma,\ee
\be \ddot\alpha + \dot \alpha^2 + \ddot\gamma + \dot\gamma^2 + \dot\alpha\dot\gamma - e^{-2\alpha} = -\bar p + 2\eta\dot\beta,\ee
\be 2 \dot\alpha - \dot \beta - \dot \gamma = -q_1,\ee
where the dot indicates time differentiation. The expansion $(\theta)$ and shear $(\sigma)$ scalars have the usual definitions \cite{14} and can be obtained using the metric (1) in the comoving coordinate system as

\be \theta = v^{\alpha}_{~\alpha} = \dot \alpha + \dot \beta + \dot\gamma,\ee
\be \sigma ^2 = {1\over 2}\sigma_{\mu\nu}\sigma^{\mu\nu} = {1\over 3}\theta^2  - (\dot\alpha\dot \beta + \dot\beta\dot\gamma + \dot\gamma\dot\alpha),\ee
where,

\be \sigma_{\mu\nu} = {1\over 2}v_{(\mu,\nu)} + {1\over 2} (v_{\mu;\alpha} v^\alpha v_{\nu} + v_{\nu;\alpha} v^\alpha v_{\mu} -{1\over 3}\theta(g_{\mu\nu} + v_\mu v_\nu).\ee
Further, as a consequence of Bianchi identity, we have

\be \dot \rho = - (\rho +p)\theta + \zeta\theta^2 + 4\eta\sigma^2 - {2\over 3} q_1\left({1\over 3}\theta^2 - \sigma^2 - \rho\right).\ee
Note that in view of equations (6) and (12), one obtains

\be {1\over 3}\theta^2 - \sigma ^2 - \rho = 3 e^{-2\alpha}.\ee
which is positive definite. Now under a suitable combination of equations (7) - (9) along with the expression of $\theta$ presented in equation (11) one finds,

\be  {\ddot\alpha - \ddot\beta\over \dot\alpha - \dot\beta} = {\ddot\beta - \ddot \gamma \over \dot\beta - \dot \gamma} = - { \ddot\gamma - \ddot\alpha \over \dot \gamma - \dot\alpha} =  = -(\theta + 2\eta) = -(\dot\alpha + \dot\beta + \dot \gamma + 2\eta),\ee
Which further on integration gives,

\be\gamma = {k\alpha + \beta \over k + 1} + \ln k_1,\ee
$k$ and $k_1$ are integration constants, whose positive magnitudes have been considered here. Additionally, in view of equations (10) and (17), one obtains,

\be q_1 = {k + 2\over k + 1} (\dot\beta -\dot\alpha),\ee
It is clearly seen from the above equation that for $\dot\alpha = \dot\beta, ~ q_1= 0$, which in view of (10) implies $\dot\alpha = \dot\beta = \dot\gamma$. This result, in turn, yields $\sigma^2 = 0$ from (11) and (12), and the metric is isotropic. On the contrary, if $q_l \ne 0$ in view of (16) and (18), one finds

\be \theta + 2\eta= - {\dot q_1\over q_1}.\ee
Further, combination of equations (11) and (17) the expression for the expansion scalar $\theta$ is found as,

\be \theta = {2k + 1\over k + 1}\dot \alpha + {k + 2\over k + 1}\dot \beta,\ee
while the expression for the shear $\sigma^2$ may be obtained in view of equations (12) and (20) as

\be \sigma^2 = {k^2 + k + 1\over 3(k + 1)^2} (\dot \alpha - \dot\beta)^2 = {k^2 + k + 1\over 3(k + 2)^2} q_1^2.\ee
In the above, we have used equation (18), which reveals that the heat flow is directly proportional to the shear scalar, i.e., $\sigma^2 \propto q_1^2$, which is truly interesting. Now the Raychaudhuri equation \cite{14} can be written as,

\be \dot\theta = -{1\over 3}\theta^2 - 2\sigma^2 + R_{\mu\nu} v^\mu v^\nu,\ee
where,

\be R_{\mu\nu}v^\mu v^\nu = -{1\over 2}[\rho + 3(p - \zeta\theta].\ee
From the above, one can note that the Hawking-Penrose energy condition is satisfied, provided $R_{ij}v^iv^j \le 0$, which implies $\rho + 3(p-\zeta\theta) \ge 0$. The energy condition, however, is always satisfied in a contracting model $(\theta < 0)$ as long as $\rho$ and $p$ are positive. The energy condition is also satisfied in an expanding model if $\rho + 3p \ge 3\zeta\theta$. Further, if the energy condition is satisfied, one can observe from (22) that $\dot\theta < 0$, which implies that a singularity is unavoidable. Now, following Belinskii and Khalatnikov \cite{15} we can write,

\be {\dot s\over s} = -{\dot \rho\over \rho + p},\ee
where, $s$ stands for the entropy density. Since the total entropy is $\Sigma = R^3 s$, so using (14) and (24) we can calculate

\be {\dot \Sigma\over \Sigma} = {\zeta\theta^2 + 4\eta \sigma^2\over \rho + p} - {2\over3 q_1}\left({{1\over 3}\theta^2 - \sigma^2 - \rho\over \rho + p}\right).\ee
The first term in the right-hand side of the above equation implies that entropy is generated by a viscous fluid and the second term says that it is generated by heat flux. The law of increase of entropy implies that the right-hand side of (25) must be greater than zero.

\section{Solution of Einstein's equations:}

Now in effect, we are dealing with eight unknown functions $\alpha,~\beta,~,\gamma,~ \rho,~ p,~ \eta,~ \zeta ~ \mathrm{and}~ q_1$ - having only five field equations, (15), (7)-(10). In order to obtain exact solutions of the field equations, we assume three physically motivated restrictions, two of which are a pair of linear relations connecting $\rho,~\sigma^2,~ \mathrm{and} ~\theta^2$, and the third being a barotropic equation of state. These are

\be \sigma^2 = C^2\theta^2,~~~\rho = D^2 \theta^2, ~~~p = \epsilon\rho.\ee
In the above $C$ and $D$ are a pair of arbitrary constants and $\epsilon$ is the polytropic index such that $0 \le \epsilon \le 1$. So now, in view of equation (26), equation (15) takes the following form,

\be \left({1\over 3} -C^2 - D^2\right)\theta^2 = 3 e^{-2\alpha}.\ee
Again, in view of equations (20), (21), and (26), we find

\be \dot \beta = h\dot \alpha.\ee
where we write $h$ for $(l \pm C m)\over (l \mp  C n)$. So now $\theta$ can be obtained from (20) and from using the above equation as

\be \theta = (m + n h) \dot\alpha.\ee
Constants $m,~ n,~ \mathrm{and}~ l$, which appear above, are all related to the constant $k$ obtained previously in the following manner,

\be m = \frac{2k + 1}{k + 1},~~~n = {k + 2\over k + 1},~~~l^2 = {k^2 + k + 1\over 3(k + 1)^2},\ee
where $m, ~n, ~\mathrm{and} ~l^2$ are all positive, since $k$ is positive in the cases under consideration and $h$ may be greater than, equal to, or less than zero. We further consider only the positive magnitudes of the constant $(m + n h)$ appearing above in the expression for $\theta$, so that the model will be expanding $(\theta > 0)$ if $\dot\alpha > 0$ and collapsing $(\theta < 0)$, if $\dot\alpha < 0$. It is to be noted that our assumed positivity of $(m + n h)$ holds for $h > 0$ and $h = 0$. It also holds for $h < 0$, provided, ${m\over n} = {(2k+ 1)\over k+2} > |h|$. Now, (27) can be written as

\be \begin{split}&{1\over \sqrt 3}\sqrt{\left({1\over 3} - C^2 - D^2\right)}~~ e^\alpha~\theta = \delta,~~\mathrm{or},\\&
{1\over \sqrt 3}( m + n h)\sqrt{\left({1\over 3} - C^2 - D^2\right)} ~~\dot \alpha e^\alpha = \delta,\end{split}\ee
where the last equation is arrived at in view of equation (29), with $\delta = \pm 1$. Either of the above pair of equations implies that the present model allows both the cases of expansion (when $\delta = +1$) and contraction (when $\delta = -1$). In what follows we only consider the case of expansion in detail and make a few remarks about contraction later.

\subsection{Expanding model:}

Integration of the equation (31) gives

\be e^\alpha = \left[{\sqrt 3 \over (m + n h)\sqrt{\left({1\over 3} - C^2 - D^2\right)}}\right](t - t_0),\ee
where $t_0$ is the constant of integration. Now $\theta$ can be obtained using equations (31) and (32) as

\be \theta = {m + n h\over t - t_0}.\ee
Since we have considered only the positive magnitude of $(m + n h)$, so the expansion implies $t > t_0$. Now (28) and (32) can be used to solve for $\beta$ as

\be e^\beta = Q (t - t_0)^h,\ee
where $Q$ is a constant of integration, which is positive-definite. Now it is easy to obtain the third metric coefficient $\gamma(t)$ from equations (17), and using equations (32) and (34) as

\be e^\gamma = k_1\left[Q\left\{{\sqrt 3\over( m + n h)\sqrt{\left({1\over 3} - C^2 - D^2\right)}}\right\}^k\right]^{1\over k - 1}(t - t_0)^{k + h \over k + 1}.\ee
The rest of the functions can now at once be evaluated in the following way. For example, $\sigma^2,~ \rho,~ \mathrm{and}~ p$ from (26) and (33), $\eta$ from equation (16), $\zeta$ from equations (22) and (23), and finally $q_1$ from equation (21), since the others are now known. So we find,

\be \sigma^2 = \left[{C(m + n h)\over t - t_0}\right]^2,\ee
\be \rho = \left[{D(m + n h)\over t - t_0}\right]^2,\ee
\be p = \epsilon\left[{D(m + n h)\over t - t_0}\right]^2,\ee
\be \eta = {1\over 2}\left[{1-(m + n h)\over t - t_0}\right],\ee
\be \zeta = {2\over 3}\left[{(m + n h)\left[{1\over 3} + 2C^2 + {1\over 2}(3\epsilon + 1)D^2\right]-1\over t - t_0}\right],\ee
\be q_1 = \left[{n(h - 1)\over t - t_0}\right].\ee
Now, the phenomenological expression for the heat conduction is given by

\be q_1 = -k(\delta_i^j + v_i v^j)(T_{,j} - T \dot v_j),\ee
where $k$ is the thermal conductivity and $T$ is the temperature. Since in our case only the $x$ component of heat flux is retained, so from the above equation we obtain

\be q_1 = - k T_{,1}.\ee
From the above equation (43) and equation (41) we can have the $x$ component of temperature gradient as

\be T_{,1} = {n(1 - h)\over k(t - t_0)}.\ee
The homogeneity consideration restricts the thermal conductivity as a function of time alone so that $k = k(t)$. Thus through integration of the above equation we can find the temperature distribution in the form,

\be T = \left[{n(1 - h)\over k(t - t_0)}\right] x + l_0 t,\ee
where $l_0$ appears as an integration constant which may either be an
arbitrary function of time or a constant. So now we get the complete set of solutions in the expanding model through (32), (34), (35), (37)-(41), and (45). The proper volume $R^3$, defined through (2) can now be readily obtained as

\be R^3 = k_1 Q^n\left({\sqrt 3\over ( m + n h)\sqrt{{1\over 3}-C^2 - D^2}}\right)^m  e^{2x} (t - t_0) ^{m + n h}.\ee

\subsection{Contracting model:}

In this model solutions can be obtained following the same procedure as above, and the corresponding quantities are obtained by simply
replacing $(t- t_0)$ by $(t_0 - t)$ and by the addition of a negative sign before the equations for $\theta,~ \zeta$, and $\eta$.\\

It can now at once be observed from the set of solutions that $\zeta$, and $\eta$ are proportional to $\rho^{l\over 2}$ which is consistent with the equations of state between $\eta,~ \rho$, and $\zeta$ considered by Belinskii and Khalatinkov \cite{15} in some special cases. If we now consider the physically reasonable condition that $\zeta > 0$ and $\eta > 0$ during expansion, then the constants appearing in the expression for $\zeta$, and $\eta$ in (39) and (40) will be positive, which implies that recontraction of the same model is not physically allowed.\\

In the following we discuss how the physical laws, such as the law of increase of entropy and the Hawking-Penrose energy condition, impose boundaries on the magnitudes of different constant parameters of the theory under consideration, e.g., $h,~ k$, etc. The law of increase of entropy follows from the second law of thermodynamics, and the strong energy condition, i.e., $R_{\mu\nu}v^\mu v^\nu \le 0$, discussed here, is satisfied for a physically reasonable classical matter.\\

\noindent
\textbf{Law of Increase of Entropy:}\\

\noindent
In view of (26), (25), and (14) we can write,

\be {\dot \Sigma\over \Sigma} = {\dot \rho + (\rho + p)\theta\over \rho + p} > 0,\ee
which, together with equations (34) and (38), further yields (remembering $t > t_0$ for the expansion) $m + nh > {2\over (1 + \epsilon)} > 0$ or $m + nh > 1$, since $0 \le \epsilon\le 1)$ which in view of equation (30) implies

\be h > 1 - {2\over n} = -{k\over (k + 2)}.\ee

\noindent
\textbf{Energy Condition:}\\

\noindent
In view of equation (22), $R_{\mu\nu}v^\mu  v^\nu \le 0$ leads to the condition $\dot \theta + {1\over 2} \theta^2 + 2 \sigma^2 \le 0$,
which together with equations (26), (33), and (30) yields

\be h \le 1 - {18C^2\over n(l+6C^2)}.\ee
Again choosing $h = {(l - Cm)\over(l + Cn)}$ it can be easily shown that the above inequality (49) reduces to (remembering $m + n = 3$ in view of Equation 30)

\be C \le {n \over 6l}.\ee
Further $h = {(l - Cm)\over(l + Cn)}$ gives, $C = {(l - lh)\over(m + nh)}$. So the above inequality on $C, n$, and $l$ can be written as an inequality on $h$ and $k$ by the use of (30) as

\be h \ge - \left({k\over k^2 + 2k +2}\right),\ee
which is stronger than the inequality condition (49). This relation leads to the condition

\be h + k \ge {k(k + 1)^2\over k^2 + 2k +2} > 0,\ee
independent of whether $h$ is positive or negative. Further, the limit on $h$ may thus be given in view of (50) and (51) as

\be 1 > h > - \left({k\over k^2 + 2k +2}\right).\ee
Again, since $k$ is an arbitrary positive constant, '$m$' and '$n$' given by (30), will be bounded having limits

\be  1 < n < 2, ~~~~~~m + n = 3.\ee

\section{Discussion on the nature of singularity:}

From the set of solutions obtained in the previous section, it is easy to
see that in the expanding model, at the initial epoch $t \rightarrow t_0$, the proper
volume vanishes and all other functions, e.g., $\theta,~ \sigma^2,~ \rho,~ p,~ \eta,~ \zeta, \mathrm{and} ~q_1$
diverge. But, since the temperature depends on thermal conductivity, which
in the spatially homogeneous model is a function of time alone, nothing
can be said with certainty about the temperature at this epoch until the
functional form of the thermal conductivity is known explicitly. However,
we can comment without hesitation that $T$ diverges at the initial epoch as
long as the coefficient of thermal conductivity remains finite. This is consistent
with the situation in big bang cosmology. At the final stage of
expansion $t \rightarrow \infty$, the proper volume diverges, and all other functions
vanish. As for the temperature we have $T \rightarrow l_0(t)$, which implies that the
universe will be in thermal equilibrium at the final stage of evolution. Now,
for the asymptotic nature of the metric coefficients given by equations (32),
(34) and (35), three cases will arise depending on $h > 0,~ h = 0,~ \mathrm{and} ~h < 0$, as
mentioned earlier. In all cases, however, as $(m + nh) > O$ we have $R^3$,
representing the proper volume, vanishing at $t \rightarrow t_0$ and diverging as $t \rightarrow \infty$.\\

\noindent
\textbf{Case 1, $h > 0$:}\\

\noindent
Here, in this case, ${\left(k + h\over k + 1\right)} > 0$. So in the expanding model as $t \rightarrow t_0$ at the initial epoch, all the metric coefficients approach zero giving rise to a point singularity. At the final stage of the evolution as $t \rightarrow \infty$, these coefficients all diverge.\\

\noindent
\textbf{Case 2, $h = 0$:}\\

\noindent
Here again ${\left(k + h \over k + 1\right)} > 0$ and $e^\beta$ no longer depends on time. So that as $t \rightarrow t_0$, $e^\alpha$ and $e^\gamma$ approach vanishingly small magnitudes giving rise to a line singularity, and as $t \rightarrow \infty$, $e^\alpha$ and $e^\gamma$ both diverge.\\

\noindent
\textbf{Case 3, $h < 0$:}

\noindent
It has been already shown in (52) that $(k + h) > 0$, even if $h$ assumes negative values. So in these cases, as $t \rightarrow t_0$, we have $e^\alpha \rightarrow 0,~ e^\beta \rightarrow \infty$ and $e^\gamma \rightarrow 0$ giving rise to a line singularity. On the other hand, as $t$ approaches infinitely large magnitude, we have $e^\beta \rightarrow 0$ but $e^\alpha \rightarrow \infty$ and $e^\gamma \rightarrow \infty$.

\section{Concluding remarks:}

In this paper, we have presented a spatially homogeneous irrotational Bianchi V cosmological model with viscous fluid and heat flux. Exact solutions are obtained with the assumptions $\sigma^2 = C^2 \theta^2,~ \rho = D^2\theta^2,~ \mathrm{and}~ p = \epsilon \rho$. The viscosity coefficients $\eta$ and $\zeta$ are found to be proportional to $\sqrt\rho$. At the initial epoch, the proper volume vanishes while all other parameters, including heat flow vector and temperature, diverge. At the final stage of evolution the proper volume takes infinitely large magnitude while other parameters such as density, pressure, shear, etc., become insignificant. So effectively, it is not required to consider isotropy as an initial condition in our model. We have started from an anisotropic universe and have shown how along with the evolution of the universe isotropization takes place due to the presence of dissipative phenomena. We have also shown that at the final stage of the evolution, temperature $T=l_0(t)$ becomes spatially homogeneous, which implies that the universe attains thermal equilibrium everywhere. We further observed that recontraction of the model is not physically allowed. Finally, since $h < 1$, as obtained in condition (49), heat flow vector $q_1$, given by expression (41), becomes negative, which further makes the second term of equation (25) a negative quantity. So we find that along with the viscosity, heat flux further adds to the rate of increase of the total entropy of the universe.

\end{document}